\documentclass[twocolumn,floatfix,showpacs]{revtex4}
\usepackage[dvips]{graphicx}
\usepackage{amsmath,amssymb}
\begin{document}
\preprint{123XXXX}
\title{\textbf{Statistical distributions of quasi-optimal paths in the traveling salesman problem:\\ 
the role of the initial distribution of cities.}}
\author{H. Hern\'andez-Salda\~na}
\email{hhs@correo.azc.uam.mx }
\affiliation{Departamento de Ciencias B\'asicas,\\
Universidad Aut\'{o}noma  Metropolitana, Azcapotzalco,\\
Av. San Pablo 180, M\'{e}xico 02200 D.F., Mexico.}
%\author{M. G\'omez-Quezada}
%\affiliation{Departamento de F\'isica, Matem\'aticas e Ingenier\'ia, Universidad de Sonora, C.P. 83600 H. Caborca, Sonora, M\'exico.}
\author{M. G\'omez-Quezada and E. Hern\'andez-Zapata}
\affiliation{Departamento de F\'isica, Matem\'aticas e Ingenier\'ia, Universidad de Sonora, C.P. 83600 H. Caborca, Sonora, M\'exico}
\date{Xxx}
\begin{abstract}
Solutions to Traveling Salesman Problem have been obtained with several
algorithms. However, few of them have discussed about the statistical
distribution of lengths for the quasi-optimal path obtained.
For a random set of cities such a distribution follows a rank 2 daisy
model but our analysis on actual distribution of cities does not show
the characteristic quadratic growth of this daisy model. The role played
by the initial city distribution is explored in this work.

\bigskip

\noindent {\it Keywords:} Traveling Salesman Problem, city distribution, statistical properties.
%The importance of understanding such a behavior in the context of
%electoral processes is discussed. 
\bigskip

Las soluciones al Problema del Agente Viajero  han sido obtenidas con varios algoritmos. Sin embargo, 
poco se ha discutido sobre la distribuci\'on de longitudes para el camino cuasi-optimal obtenido. 
Para un conjunto al azar de ciudades esta distribuci\'on sigue un modelo margarita de rango 2, pero
nuestro an\'alisis sobre distribuciones reales de ciudades no muestra el crecimientos cuadr\'atico 
caracter\'istico de este modelo margarita. En este trabajo se explora el rol jugado por la distribuci\'on 
inicial de ciudades. 

\bigskip
\noindent {\it Descriptores:} Problema del Agente Viajero, distribuci\'on de ciudades, propiedades estad\'isticas.
\end{abstract}


\maketitle

\section{Introduction}

Statistical approaches to complex problems have been successful 
in a wide range of areas, from complex nuclei \cite{Wigner} to the 
statistical program for complex systems (see for instance Ref. 
\cite{Palis}), including the fruitful analogies with
statistical mechanics. The main goal is to find universal properties,
i.e., properties that do not depend on the specifics of the system treated
but on very few symmetry or general considerations. An example of such 
an approach is represented by the Random Matrix Theory (RMT) \cite{Mehta3ed,WeidenmullerGuhr}
which has been applied successfully to wave systems in a range of 
typical lengths from one femtometre to one metre. 

Attempts to apply a RMT approach to a non-polynomial problem like the Euclidean Traveling Salesman
problem (TSP) is not the exception \cite{Flores,hhs}. For an {\it ensemble} of cities, randomly 
distributed, general statistical properties appear and they are well described for a
daisy model of rank $2$\cite{hhs}. 
However, in realistic 
situations the specific system seems to be important, as we shall see later, and 
the initial conditions rule the statistical properties of the solutions. 
In the present work we deal with a similar analysis both for actual TSP maps for several 
countries through the 
planet and for some toy models that could help to understand the role of 
initial conditions in the transition to universal statistical properties. 
An interesting link appears with the distribution of corporate vote when we map 
the lengths in the TSP problem to the number of votes, after a proper normalization.

The paper is organized as follows: In section II we define the TSP
and the statistical measures we shall use. 
Special attention will be paid to 
the separation of secular and fluctuation properties. Here, we analyze  maps of 
actual countries. 
In section III we discuss the transition from a map defined on a rectangular grid 
to a randomly distributed one using two random perturbations to the city positions: i) a 
uniform random distribution with a width $\sigma$ and, ii) a Gaussian distribution.
In the same section, we discuss the relation of the latter toy model to the distributions
of corporate vote. The conclusions appear in section IV.

\section{Statistical properties of quasi-optimal lengths}

In the Traveling Salesman Problem  
a seller visits $N$ cities  with given positions $(x_i,y_i)$, returning
to her or his city of origin. Each city is visited only once and the task is to 
make the circuit as short as possible. This problem pertains to the category of so called 
{\it NP-complete} problems, whose computational time for an exact solution increases
as an exponential function  of $N$. It is, also, a minimization problem  and
it has the property that the  objective function has many local minima~\cite{NumericalRecipes}.

Several algorithms exist in order to solve it and the development of much more efficient ones is matter 
of current research.
Since we are interested in statistical properties of the quasi-optimal paths, small differences between the different 
algorithms are not of relevance and we shall consider all of them as equivalent. 
The computational time is irrelevant as well.
In this paper we used the results of Concorde~\cite{Concorde} for the actual-country TSP, and simulated 
annealing~\cite{NumericalRecipes} and 2-optimal~\cite{Croes} for the  analysis of specific models.

The first step in the analysis consist in separating the fluctuating properties from 
the secular ones in the data. This process could be nontrivial \cite{WeidenmullerGuhr}.
The idea behind this procedure is that all the peculiarities of the system resides in 
the secular part and the information carried out by the fluctuations have an universal character. 
In the energy spectrum of many quantum systems, when almost all the symmetries are broken,
this kind of analysis has shown that the fluctuations are universal and regarding 
only to the existence or not of a global symmetry like time reversal invariance. 
The peculiarities of the system, wheather it is a many particle nucleus, an 
atom in the presence of a strong electromagnetic field, or a billiard with a chaotic  classical dynamics,
all these characteristics are in the secular part~\cite{WeidenmullerGuhr}. In the present case, we assume that the 
dynamics that rules where the cities are located  is sufficiently complex in order to admit
this kind of analysis. If not, as we shall see later, the next step in our analysis is to search 
for the reason of such a lack of universality.

In order to perform this separation  we consider the density of cumulative lengths, $d$, named
\[
\rho(d) = \sum_k \delta(d-d_k),
\]
with $\delta(\cdot)$, the Dirac's delta function. The cumulative lengths $d_k$ are ordered as they appear in the 
quasi optimal path and are defined as $d_k=\sum_{i=1}^k l_i $ with 
$l_i=\sqrt{(x_i-x_{i-1})^2+(y_i-y_{i-1})^2}$ being the length between city $i-1$ and $i$.
The cities are located at $(x_i,y_i)$ in the $X-Y$ plane. The corresponding cumulative density  is 
\[
{\cal N}(d) = \sum_k \Theta(d-d_k),
\]
with $\Theta(\cdot)$, the Heaviside function. The task is to separate ${\cal N}(d)$ as
${\cal N}_{Secular}(d)+{\cal N}_{fluctuations}(d)$. 
The secular part is calculated using a polynomial fitting of degree $n$. After this, 
we consider as the variable to analyze the one  mapped as 
\[
\xi_k = {\cal N}_{Secular}(d_k).
\]

We shall study the distributions of the set of numbers ${\xi_k}$. 
Notice that this transformation makes that $\langle \xi \rangle = 1$.
The analysis is performed on windows of different size, this kind of analysis is 
always of local character. For historical reasons this spreading procedure is named unfolding.
From all the statistics we shall focus on the 
nearest neighbor distribution, $P(s)$, with $s_i=\xi_i-\xi_{i-1}$ (which is the normalized length), 
and the number variance, $\Sigma^2(L)$,
for the short and large range correlations, respectively. The $\Sigma^2(L)$ is 
the variance in the number of levels $\xi_n$ in a box of size $L$, see \cite{Mehta3ed,WeidenmullerGuhr}
for a larger explanation.

The actual maps considered in this work were those which are reported in Concorde's web page
\cite{Concorde}, and we selected those that have a number of cities larger than 
$1000$ and present no duplications. The countries selected are reported in table \ref{Tab:1}.

\begin{table}
\begin{tabular}{llrrr}
\hline
 & Name   &  N cities & $l_{max}$   &  $<l>$           \\
\hline
1 & Burma  &  33708    &  8650.00  & 197.00       \\ %  bm33708
2 &Canada &  4663     & 76133.33  & 10263.28     \\ % ca4663.nde
3 & China  & 71009     &  51646.00  & 4225.62     \\ %ch71009
4 &Egypt  & 7146      &  6708.34 & 431.90    \\ %eg7146
5 & Ireland & 8246   & 4059.17  & 973.62    \\ %  ei8246
6 &Finland & 10639     & 10200.01 & 536.04    \\ %  fil10639
7 &Greece  & 9882   &   6600.02   & 286.97   \\ %gr9882
%Honduras & 14473   &  5836.76    &  1095.80  \\ %ho14473 w repps
8 & Italy   & 16862  & 10050.00  &  438.46   \\ % it16862
9 &Japan  & 9847 &  16166.68 & 581.44    \\ %ja9847
10 & Kazakhstan & 9976 & 38925.00 & 5559.56     \\ % kz9976
11 & Morocco &  14185 & 8056.41  & 418.08   \\ %  mo14185
12 & Oman    & 1979   & 3685.56 & 541.17 \\ %mu1979
13 & Sweden  & 24978  & 9250.02 & 368.03 \\ %sw24978
14   & Tanzania  & 6117 & 10166.68 &  1145.41 \\ %tz6117
15   & Vietnam & 22775  & 4983.36   & 145.26 \\ %vm22775
16    & Yemen & 7663    & 10810.28 & 877.26 \\ % ym7663
\hline
\end{tabular}
\caption{\label{Tab:1} Countries considered for the study. The quasi-optimal
path was obtained from the web page in Ref.~\cite{Concorde}. Maximum and average length (in {\it km})
are reported too. }
\end{table}

%A plot of 
%${\cal N}(l)$ appear in Fig. ~\ref{Fig:1}. In (a) appear the results for all the cases 
%of Table \ref{Tab:1} except of China and Holland, which appear in (b) and (c), respectively.  
%As can see, the variations are large, from highly fluctuating countries like China to 
%smooth ones like Holland. 

\begin{figure}[tb]
\includegraphics[width=\columnwidth,bb= -29 -12 764 583 ]{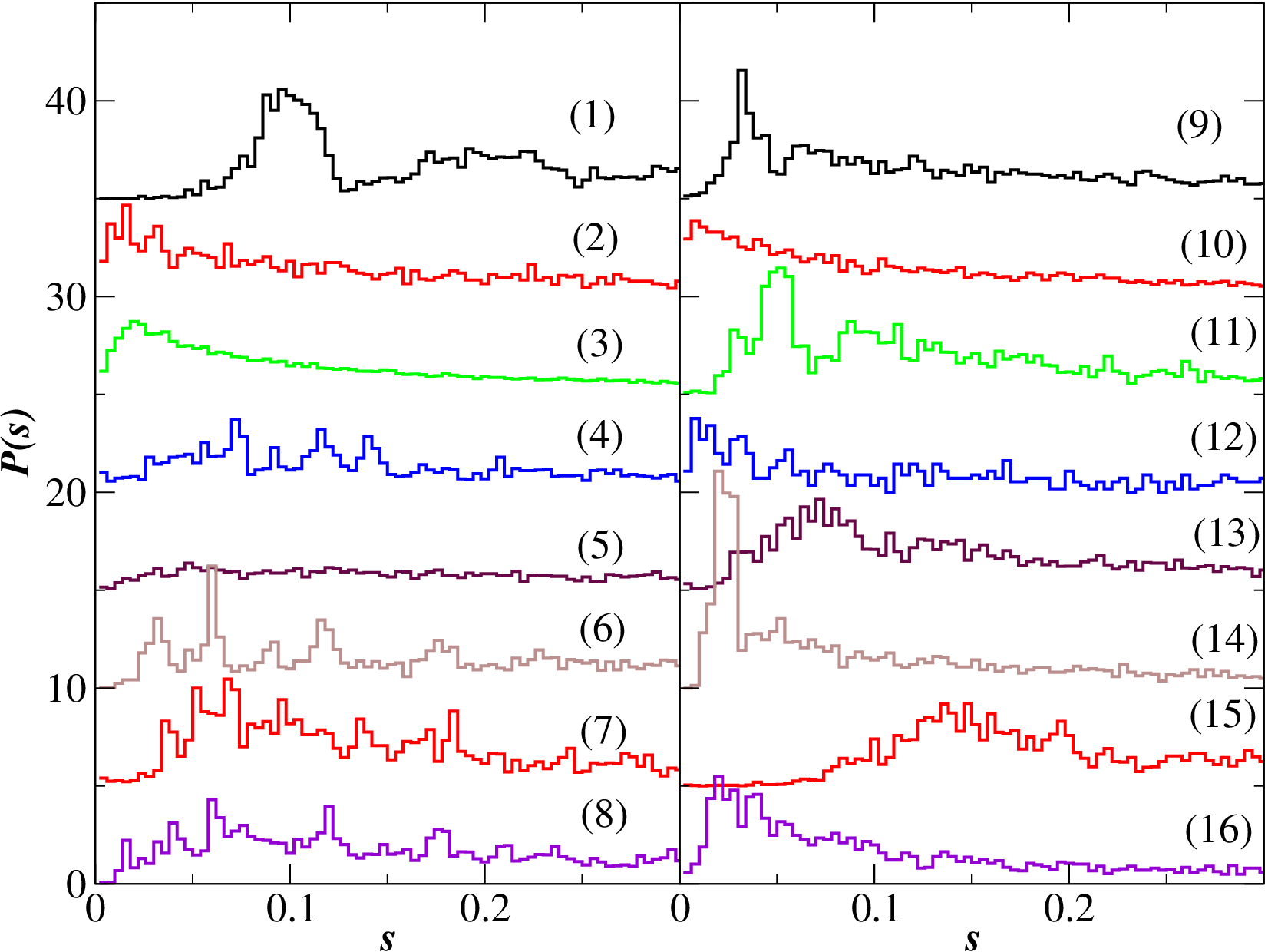}
\caption{(color online) Nearest neighbor distribution of normalized lengths (histogram), for the
quasi-optimal path in the countries referred in Table \ref{Tab:1}. For sake of  clarity, we shifted the distributions. 
The bin size is $0.004$ and notice that the normalized average is $s=1$.}
\label{Fig:0}
\end{figure}

\begin{figure}[tb]
\includegraphics[width=\columnwidth, bb = -46 -12 769 583]{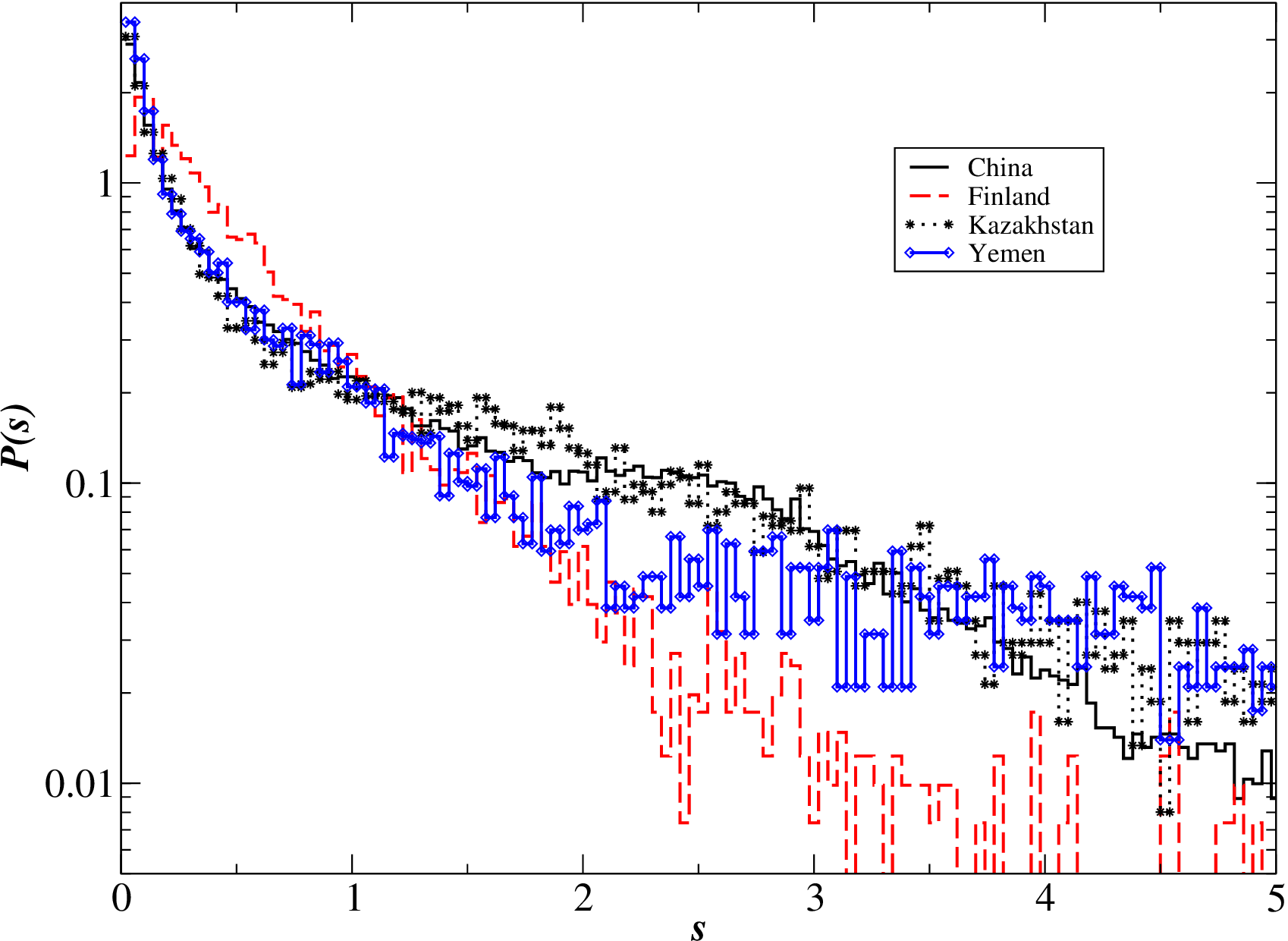}
\caption{(color online) Nearest neighbor distribution of normalized lengths in semi-log scale, for the
quasi-optimal path in the countries referred in the inset. The bin size is $0.04$.}
\label{Fig:1}
\end{figure}

The results in the ${\cal N}(l)$ are full of variations, as expected. 
The secular part was calculated with several windows size and polynomial 
degrees looking for those parameter values that stabilize the statistics. However, 
a no universal behavior appears. 
The graphs presented in Fig. \ref{Fig:0} for the nearest neighbor distribution
were calculated using a polynomial fitting of fourth degree and windows 
of $200$ lengths, the histograms have a bin of size $0.004$. There we show the distribution for all the countries
in table \ref{Tab:1}. 
There is no a single distribution 
even when some of the countries show a exponential decay as occurs  with Finland (see Fig. \ref{Fig:1}). 
We analyzed the data with several windows size and polynomial degrees with similar results.

No regularities were found in our analysis but all the histograms show almost a maximum and some of them 
present a polynomial grow, $s^{\alpha}$, at the beginning of the distribution. This could be seen in Burma (1),
Japan (2) and Sweden (13). Notice that the distributions present the maximum at $s < 0.2$ meanwhile the 
average is at $s=1$. Hence, all the distributions present a long tail. However the type of decay is diverse 
as well (see Fig. \ref{Fig:1}). Some distributions present a clear exponential decay, like Finland. Some others present a mixed decay 
as is the case of China. For sake of clarity we do not show all the cases in Fig. \ref{Fig:1}. Notice 
that the bin size is larger in Fig. \ref{Fig:1} compared to that used in 
Fig. \ref{Fig:0}, for this reason, the polynomial grow does not appear.

We try to show countries of several sizes and urban configurations. There are some countries 
with an urban density almost constant, like Sweden, and some others with long tails and a maximum
for very short distances as Canada. Changes in the parameters of analysis, windows size and fitting 
polynomial degree,
do not give us an universal behavior. Recall that this is not the case for an {\it ensemble} 
of randomly distributed cities as seen in Ref.\cite{Flores,hhs}.
From all these results it is clear that the  initial distribution of  cities plays 
a crucial role in the final quasi-optimal result. This is the subject of the next 
section.

\section{The role of the initial distribution of cities  }

\begin{figure}[tb]
\includegraphics[width=\columnwidth, bb = -1 11 763 583]{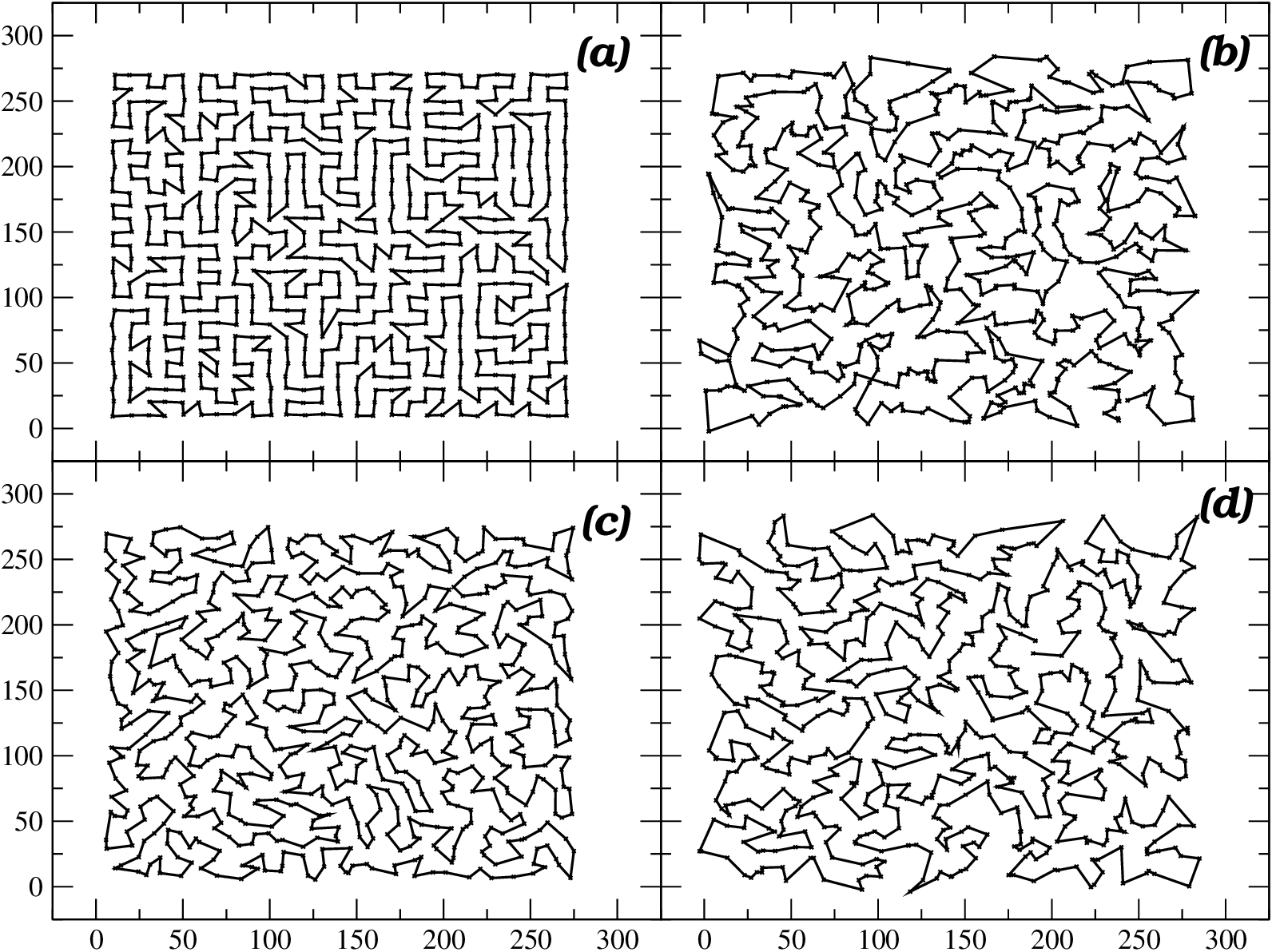}
\caption{ Examples of the quasi-optimal path on one map of the model I, with 
values of (\textrm{a})$\sigma =1.0 $ and (b)$\sigma=15.0$. For model II, the considered values are 
(c) $\sigma=5.0$ and (d) $\sigma = 10.0$. The maps  consist of $27 \times 27$ cities. 
}
\label{Fig:2a}
\end{figure}
The statistical properties of quasi-optimal paths for an ensemble of  randomly distributed
cities in the Euclidean plane is well described by the so  called daisy model of 
rank $r = 2$~\cite{hhs}. 
These models are the result of retaining each $r+1$ number from a sequence of random numbers $y_i$
which follow a Poisson distribution, i.e, its $n$-nearest neighbor distribution has the form 
\[
P(n,s)= \frac{s^{n-1}}{(n-1)!}\exp(-s),
\]
with $s_i= y_{i+n}-y_i$ and $n=1$ corresponds to first nearest neighbors. The rarefied sequence must be re-scaled in order to recover the norm and
the proper average of the $n$-nearest neighbor distributions.

For the general daisy model of rank $r$ we have the following expression for the nearest neighbor distribution:
\begin{equation}
P_r(s) = \frac{(r+1)^{(r+1)}}{\Gamma(r+1)} s^r \exp( -(r+1)  s),
\label{pdsdaisy}
\end{equation}
and, 
\begin{equation}
\begin{split}
\Sigma^2_r(L) =  \frac{L}{r+1} + \frac{r(r+2)}{6(r+1)^2} +\frac{2}{(r+1)^2} \sum^r_{j=1} 
\frac{W_j}{(1-W_j)^2} \\ 
\times \exp \left[ (W_j-1)(r+1) L)\right],
\end{split}
\label{s2daisy}
\end{equation}
for the number variance. Here $W_j = \exp(2\pi i j/(r+1))$ are the $r+1$ roots of unity and $i$ stands for $\sqrt{-1}$.

In the case of $r=2$, both, the nearest
neighbor distribution and the $\Sigma^2_2$ statistics have the theoretical results,
namely
\begin{equation}
P_2(s) = \frac{3^3}{2!} s^2 \exp(-3 s),
\label{pdsdaisy2}
\end{equation}
and
\begin{equation}
\Sigma^2_2(L) =  \frac{L}{3} + \frac{4}{27} \left[ 1- \cos(\frac{3\sqrt{3}}{2}) \exp(-\frac{9}2{}L)\right].
\label{s2daisy2}
\end{equation}
As mentioned above, quasi-optimal paths of an {\it ensemble} of maps with cities randomly distributed nearly follow 
equations (\ref{pdsdaisy2}) and (\ref{s2daisy2}). Again, the final distribution of lengths in the quasi-optimal paths
depends on the initial distribution of cities.

\begin{figure}[tb]
\includegraphics[width=\columnwidth, bb = 16 54 705 523 ]{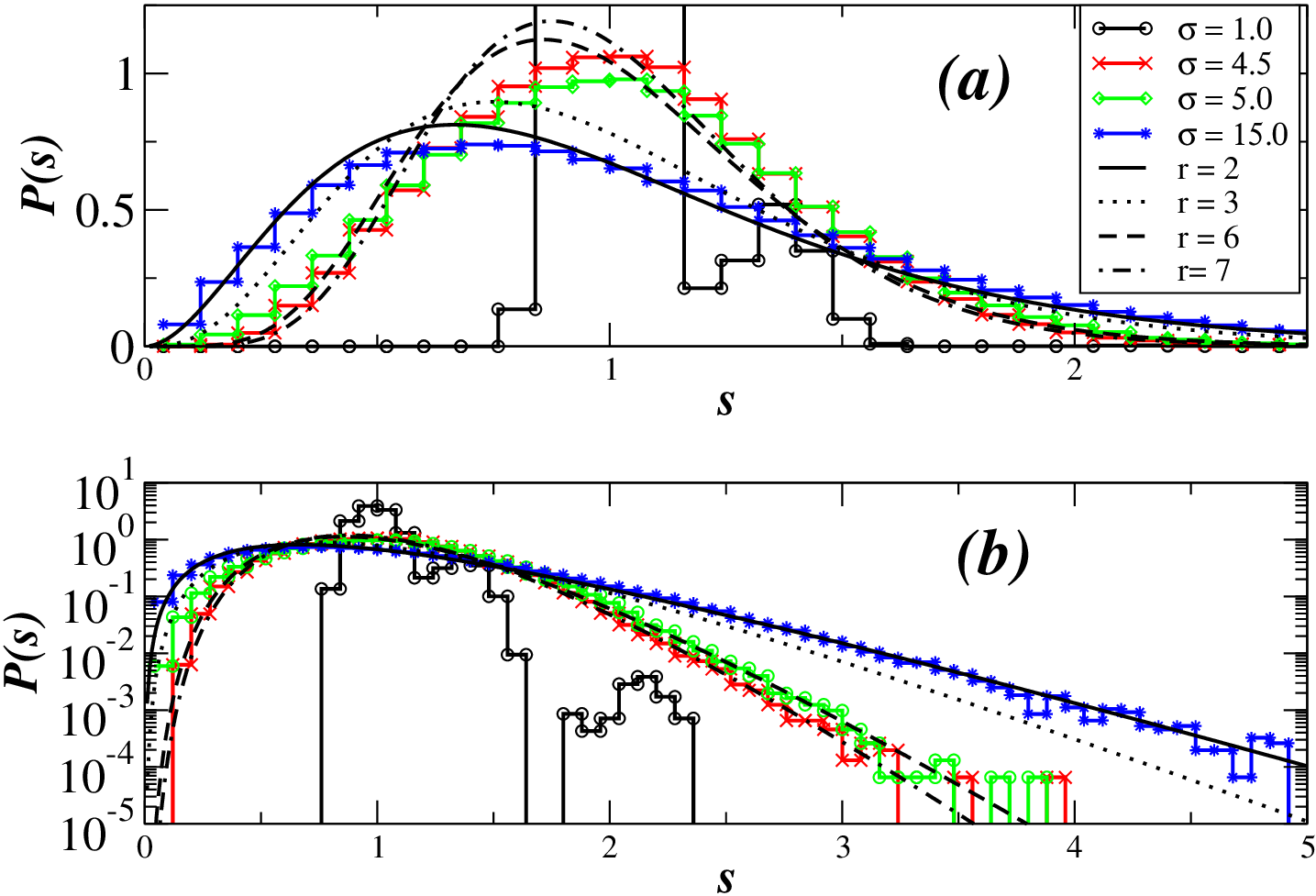}
\caption{(color online) (a) Nearest neighbor distribution, $P(s)$, of normalized lengths 
for model I for the value of the $\sigma$ indicated in the key.  
The functions in continuous line and labeled with $r$ correspond to the distributions 
of the daisy model of equation \ref{pdsdaisy}. (b) Same as (a) but in semi-log scale.
Note that the histogram for $\sigma = b/2 = 5.0$
is well fitted by the daisy model with $r=6$.(See text for details)}
\label{Fig:2}
\end{figure}

In order to understand the role of initial distribution of cities, 
we depart from a master map of cities on a square grid of side $b$, where the initial position 
of each city is in the intersections. In this case, the distribution of lengths for the quasi-optimal path
is close to a delta function 
with a small tail. Now, we take an ensemble of maps, each one is built up  
relocating the cities from their original positions using a probability distribution of 
width $\sigma$, ${\cal P}(\sigma)$. That is 
\begin{eqnarray}
x_i = n b + p_i \\
y_i = m b + q_i,
\end{eqnarray}
where the numbers $p_i$ and $q_i$ are taken from ${\cal P}(\sigma)$ which have
zero mean and $n$ and $m$ are integers. 
Two distributions were selected, i) a uniform one  with width $\sigma$, that we call model I, and ii) a 
Gaussian one with the  same width, that we named model II.  
In Figure \ref{Fig:2a} several examples are shown of the quasi-optimal paths. The starting grid or 
master rectangle defining the $27\times27$ cities is similar to that of Figure \ref{Fig:2a}(a).

\begin{figure}[tb]
\includegraphics[width=\columnwidth, bb = 19 50 707 529]{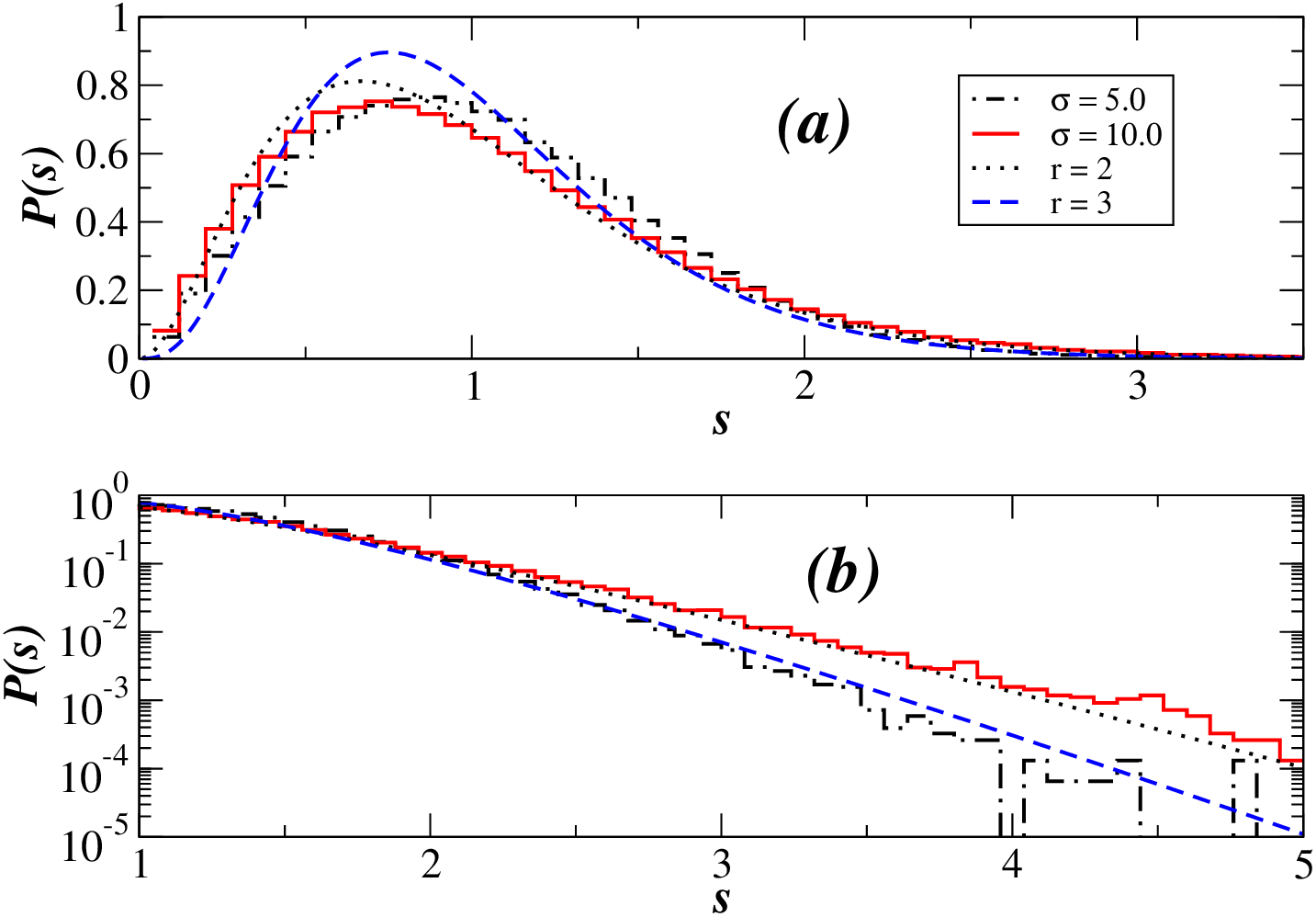}
\caption{(color online) In the same order as previous figure, but for
model II and the values of $\sigma$ and $r$ indicated. Notice that this time 
the value $\sigma = b/2 =5.0$ is fitted by a daisy model or rank 
$r=3$.}
\label{Fig:3}
\end{figure}

The distribution of lengths shows a 
transition from  the original one to 
the limit case given by equations (\ref{pdsdaisy2}) and (\ref{s2daisy2}) as can be seen in Fig. 
\ref{Fig:2} for a uniform distribution and in Fig.~\ref{Fig:3} for the Gaussian one. 
In these figures we plot $P(s)$ in both (a) linear and (b) logarithmic scale. 
The analysis was performed on an ensemble of $500$  maps of $27\times 27$ cities each. The variable $\sigma$ was considered
in the interval $[0,15]$. Only relevant values are reported.

For model I, we plotted the histograms for $\sigma = 1, 4.5, 5$ and $15.0$ (see Fig.\ref{Fig:2}). For the first case 
the distribution departs barely from the initial one but, the quasi optimal solution 
presents a revival for $s$ slightly below $\sqrt{2}$(Fig.~\ref{Fig:2}(a) the black histogram with circles), and for values 
slightly larger than $2$ representing the existence of diagonal lengths in the grid and 
lengths of order two $b$'s (in this uncorrelated variable, see Fig. \ref{Fig:2a}(a)). The histograms show 
a continuous transition to the distribution given by Eq.~(\ref{pdsdaisy2}) when 
$\sigma=15$ (blue histogram with stars). 
The tail, in this case, follows very closely the daisy model of rank 2 to it and the start of $P(s)$ is consistent with it (see Fig. \ref{Fig:2}(b)). 
The histogram in red with crosses corresponds to $\sigma = 4.5$ and follows very closely the daisy model of rank $7$. Meanwhile, the histogram in green with diamonds, corresponding to $\sigma = 5.0$, follows the rank $6$ model.
The second case corresponds to the value of
$\sigma$ when the distribution of the cities start to admit overlapping, i.e $\sigma =b/2$. When 
$\sigma = b$ the histogram coincides with the daisy model of rank $3$ (not shown).

For model II there exists a transition and the fitting to a daisy model is, as well, defined as in model I.
In Fig. \ref{Fig:3} we plot the cases $\sigma = 5.0$ and $10.0$ which are close to daisy models of 
rank $3$ and $2$, respectively. In Fig. \ref{Fig:3}(b) we re-plotted in semi-log scale in order to see the tail decay. 
The interpretation is the same as the previous one.
Fluctuations 
in the tails for large values of  $\sigma$ are observed in the Gaussian case. The reason for 
them is that the map admits cities positioned far away  from the master rectangle (not shown). 
Certainly, a fit using Weibull or Brody distributions is possible for both models, however 
it does not exist a link with any physical model whereas the daisy models 
are related to the 1-dimensional Coulomb problem~\cite{hhs}.

An interesting link exists in this context between the 
distribution of lengths in the quasi-optimal path and 
the vote distribution trough Daisy models. 
In Ref. \cite{hhse1} it has been established that the  distribution  
of corporate vote in Mexico, during elections from $2000$ to $2006$ 
follows a daisy model with ranks from $r=2$ to $r=7$. In the present case,
exponential decays that are compatible with $r=3$ 
occur at $\sigma=b$ for model I and $\sigma=b/2$ for model II, i.e., $\sigma = 10.0$ for
the former and $\sigma= 5.0 $ for the latter.
In both cases, distributions of the random perturbation 
overlap is significant. Other links look possible and are 
under consideration.
This behavior remains poorly understood, but it 
opens new questions about the relationship between statistical
mechanics problems and social behavior.
 
In the case of long range behavior, the  analysis with 
the $\Sigma^2$ statistics looks promising, but the 
asymptotic behavior does not coincide with daisy models. 
This statistics is highly sensitive to the unfolding 
procedure described before. Wider studies are currently 
in progress, however we give in advance  that
the $\Sigma^2$ statistics for model I at $\sigma=15.0$ is close to 
that of the daisy model of rank 2. The slope is $0.317486 \pm 7.31\times 10^{-5}$ 
which is close to the $1/3$ value obtained for the daisy model. For both models the $\Sigma^2 $ starts following the behavior 
of Eq. (\ref{s2daisy2}). For small values of $\sigma$ the numerical results show
an oscillatory behavior compatible with the presented in the daisy model, Eq.~(\ref{s2daisy}),
even when the asymptotic slope is not the correct one.

\section{Conclusions}
We presented a statistical approach to the Traveling Salesman problem (TSP). 
No universal behavior appears in the case of actual distribution of 
cities for several countries world wide as it appears in the case
of the Euclidean TSP with a uniform random distribution of cities \cite{Flores,hhs}. As a first step 
to understand  the role of the initial distributions of cities, we study the 
nearest neighbor distribution for the lengths of quasi-optimal paths
for a model which start with a periodic distribution of cities on 
a grid and it is perturbed by a random fluctuation of width $\sigma$. We use two models for the 
fluctuation: model I, a uniform distribution and, model II, a Gaussian one. Both 
models evolve, as a function of $\sigma$, from a delta like initial distribution to  
one well described by a daisy model of rank $2$ (see Eq.~(\ref{pdsdaisy2})).
As the perturbative distribution width is increased the evolution of the models present a 
nearest neighbor distributions 
compatible with several ranks of the daisy model. Two values of the width are 
important, the first one is when the random perturbation admits an overlap 
of the cities originally at the periodic sites. For model I that occurs when $\sigma= b$ (the total width of
the distribution is $2\sigma$), being $b$ the distance of the periodic lattice.  
For model II that happens when $b=2\sigma$, i.e. two
standard deviations of the Gaussian distribution. In these 
cases the histogram of lengths fits a rank $3$ daisy model. An interesting
link appears when we notice that such a daisy model fits the tail of
the distribution of votes (for the chambers) for a corporate party in Mexico 
during election of 2006.  The reason of this coincidence remains open and 
requires further analysis. An attempt in this direction is presented in Ref.\cite{hhs2010}.
Another open question concerns about if an {\it ensemble} of 
world wide countries have universal properties or not. This topic is in
current research.

\section*{Acknowledgments}
HHS was supported by PROMEP 2115/35621 and partially by DGAPA/PAPIIT IN-111308.

\end{document}